\documentclass[%
 aip,
 amsmath,amssymb,
 reprint,%
]{revtex4-1}

\usepackage{graphicx}
\usepackage{dcolumn}
\usepackage{bm}
\usepackage{float}
\usepackage{subcaption}
\usepackage{hyperref}

\usepackage[utf8]{inputenc}
\usepackage[T1]{fontenc}
\usepackage{mathptmx}
\usepackage{etoolbox}

\makeatletter
\def\@email#1#2{%
 \endgroup
 \patchcmd{\titleblock@produce}
  {\frontmatter@RRAPformat}
  {\frontmatter@RRAPformat{\produce@RRAP{*#1\href{mailto:#2}{#2}}}\frontmatter@RRAPformat}
  {}{}
}%
\makeatother
\begin{document}

\preprint{AIP/123-QED}

\title{Effect of Nozzle Curvature on Supersonic Gas Jets Used in Laser-Plasma Acceleration}

\author{Ocean Zhou}
\email{oceanzhou123@berkeley.edu}
\affiliation{Lawrence Berkeley National Lab (LBNL), CA 94720, USA}
\affiliation{University of California, Berkeley, CA 94720, USA}
\author{Hai-En Tsai}
\affiliation{Lawrence Berkeley National Lab (LBNL), CA 94720, USA}
\author{Tobias M. Ostermayr}
\affiliation{Lawrence Berkeley National Lab (LBNL), CA 94720, USA}
\author{Liona Fan-Chiang}
\affiliation{Lawrence Berkeley National Lab (LBNL), CA 94720, USA}
\affiliation{University of California, Berkeley, CA 94720, USA}
\author{Jeroen van Tilborg}
\affiliation{Lawrence Berkeley National Lab (LBNL), CA 94720, USA}
\author{Carl B. Schroeder}
\affiliation{Lawrence Berkeley National Lab (LBNL), CA 94720, USA}
\affiliation{University of California, Berkeley, CA 94720, USA}
\author{Eric Esarey}
\affiliation{Lawrence Berkeley National Lab (LBNL), CA 94720, USA}
\author{Cameron G. R. Geddes}
\affiliation{Lawrence Berkeley National Lab (LBNL), CA 94720, USA}

\date{\today}

\begin{abstract}
Supersonic gas jets produced by converging-diverging nozzles are commonly used as targets for laser-plasma acceleration experiments. A major point of interest for these targets is the gas density at the region of interaction where the laser ionizes the gas plume to create a plasma, providing the acceleration structure. Tuning the density profiles at this interaction region is crucial to LPA optimization. A "flat-top" density profile is desired at the line of interaction to control laser propagation and high energy electron acceleration, while a short high-density profile is often preferred for acceleration of lower-energy tightly-focused laser plasma interactions. A particular design parameter of interest is the curvature of the nozzle's diverging section. We examine three nozzle designs with different curvatures: the concave "bell", straight conical and convex "trumpet" nozzles. We demonstrate that for mm-scale axisymmetric nozzles that, at mm-scale distances from the nozzle exit, curvature significantly impacts shock formation and the resulting gas jet density field and therefore, is an essential parameter in LPA gas jet design. We show that bell nozzle is also able to produce focused regions of gas with higher densities. We find that the trumpet nozzle, similar to the straight and bell nozzles, can produce flat-top profiles if optimized correctly and can produce flatter profiles at the cost of slightly wider edges. An optimization procedure for the trumpet nozzle is derived and compared to the straight nozzle optimization process. We present results for different nozzle designs from computational fluid dynamics simulations performed with the program ANSYS Fluent and verify them experimentally using neutral density interferometry. 
\end{abstract}

\maketitle

\section{\label{sec: Intro} Introduction}

\begin{figure}[H]
\centering
    \begin{subfigure}{0.3\textwidth}
    \centering
        \includegraphics[width = \textwidth]{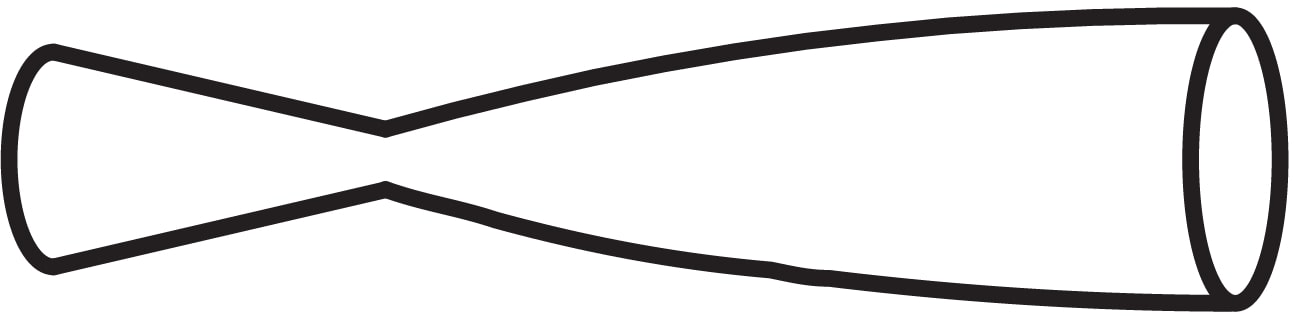}
        \caption{Concave "Bell" aka Parabolic}
        \label{fig: bell}
    \end{subfigure}
    \begin{subfigure}{0.3\textwidth}
    \centering
        \includegraphics[width = \textwidth]{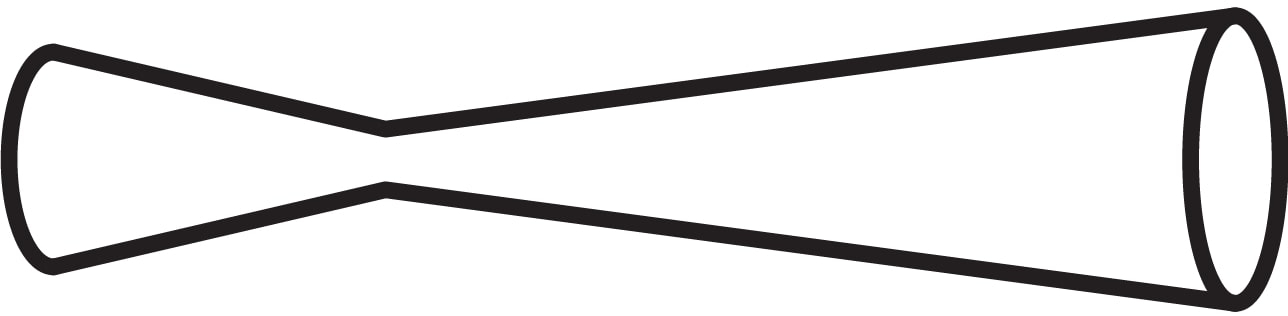}
        \caption{Straight}
        \label{fig: straight}
    \end{subfigure}
    \begin{subfigure}{0.3\textwidth}
    \centering
        \includegraphics[width = \textwidth]{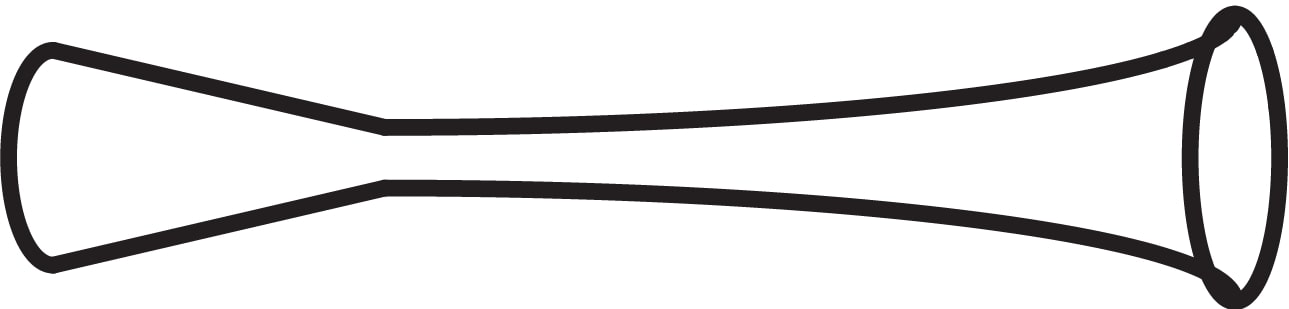}
        \caption{Convex "Trumpet"}
        \label{fig: trumpet}
    \end{subfigure}
    \caption{Three-dimensional profile of each nozzle considered in this paper}
    \label{fig: nozzleprofs}
\end{figure}

Laser-plasma acceleration (LPA) is a particle acceleration scheme that uses an ultrafast intense laser pulse to create a plasma wave that can sustain strong acceleration gradients of hundreds of GV/m to achieve electron acceleration over short distances \cite{lpatajima, lpaesarey, LPAexp, lpamalka, lpaplasmawave, LPAhighgrad, LPAwfgen, LPAnonlinear, direcgrad}. Such accelerators have become capable of producing relativistic quasi-monoenergetic electron beams \cite{krushelnick, geddesnature, faurenature, THzIFPhysLett} in the hundreds of MeV \cite{leenatphot, Kneip, Froula, pukhovmono, ControlledInjMono, attosec} to above GeV energy level \cite{LeemansNature, XWangNature, KimPhys, wimPhysLett, KimPhysLett, PW8GeV}. Controlled LPA experiments require a well-defined interaction region between the laser pulse and the plasma target \cite{tonypaper, highdensgasjet, Lemos, cgrth}. The plasma target is typically created by ionization of a gas target in the onset of high power laser pulses used to drive the plasma wave. A long flat-top density profile is often desired for laser propagation, plasma waveguide creation, and electron acceleration \cite{cgrth, Milchberg, gasjetSchmid, Semushin}. On the other hand, a sharp high density profile is useful for high repetition rate LPA driven by mJ-level laser pulses \cite{kHzLPA1, kHzLPA2} and electron injection \cite{liona, kkswan, haien, sambarber, SchmidInjection, CGeddesInjection, BuckInjection, GuillaumeInjection, ThauryInjection, VeiszInjection}. One common method of creating desired density profiles is by producing a supersonic gas jet through converging-diverging (C-D) nozzles. Numerous studies have been conducted in the context of LPA on nozzle manufacturing techniques \cite{3Dprinting} and nozzle design \cite{gasjetSchmid, gastargetsLorenz, liona, earlystudyJLHen, MKrish, cgrth, Semushin, Vmalka, highdensgasjet, Lemos, MinIOP, MusinskiGasJet, YMLi, froulaIF}. While the diverging section curvature of C-D nozzles has been examined and optimized for various applications in past studies \cite{Nasatrumpet, Eriksson, MKrish, Atkinson, windtunnel, DengThesis}, more investigation of the diverging section curvature's effect in an LPA context is needed.

In this paper, we examine three axisymmetric mm-scale nozzle designs with different curvatures shown in Fig. \ref{fig: nozzleprofs}(a), (b), (c). Although non-axisymmetric nozzles, such as rectangular slit nozzles, have also been used to produce gas jet targets \cite{gastargetsLorenz}, axisymmetric nozzles were chosen in order to keep computation and manufacturing simple. The trumpet design, which has not been commonly studied or applied in an LPA context, is based off of previous designs in other fields \cite{ogtrump, Nasatrumpet}. Simulation results suggest that the nozzle curvature has great effect on shock formation and the resulting density field outside the nozzle exit. It is found that the trumpet nozzle, like the straight and bell nozzles, can effectively yield flat-top density profiles. Optimization of the curvature of such nozzles can assist in the formation of very flat profiles. Furthermore, it is also found that the bell curvature can create highly focused regions of gas with large density fluctuations, expanding the bell nozzle's applications beyond creating flat-top profiles \cite{Lemos}. Finally, an optimization procedure is developed for the trumpet nozzle to create desired flat-top profiles.

 To experimentally verify simulation results, we employ neutral density interferometry, a popular gas jet characterization method \cite{earlystudyJLHen, Vmalka, gasjetSchmid, gastargetsLorenz, Semushin, highdensgasjet, MKrish, Lemos, MusinskiGasJet, YMLi, froulaIF}. The paper is structured as follows: Sec. \ref{sec: Sims} describes the principles of nozzle design and the simulation methods used. Sec. \ref{sec: Experiment Method} presents the diagnostic setup. Sec. \ref{sec: Results} presents simulation results and the comparison to measurements and Sec. \ref{sec: discon} summarizes the study and discusses future areas of interest.

\section{\label{sec: Sims} Gas Jet Simulations}
\begin{figure}[H]
    \centering
    \includegraphics[width = 0.5\textwidth]{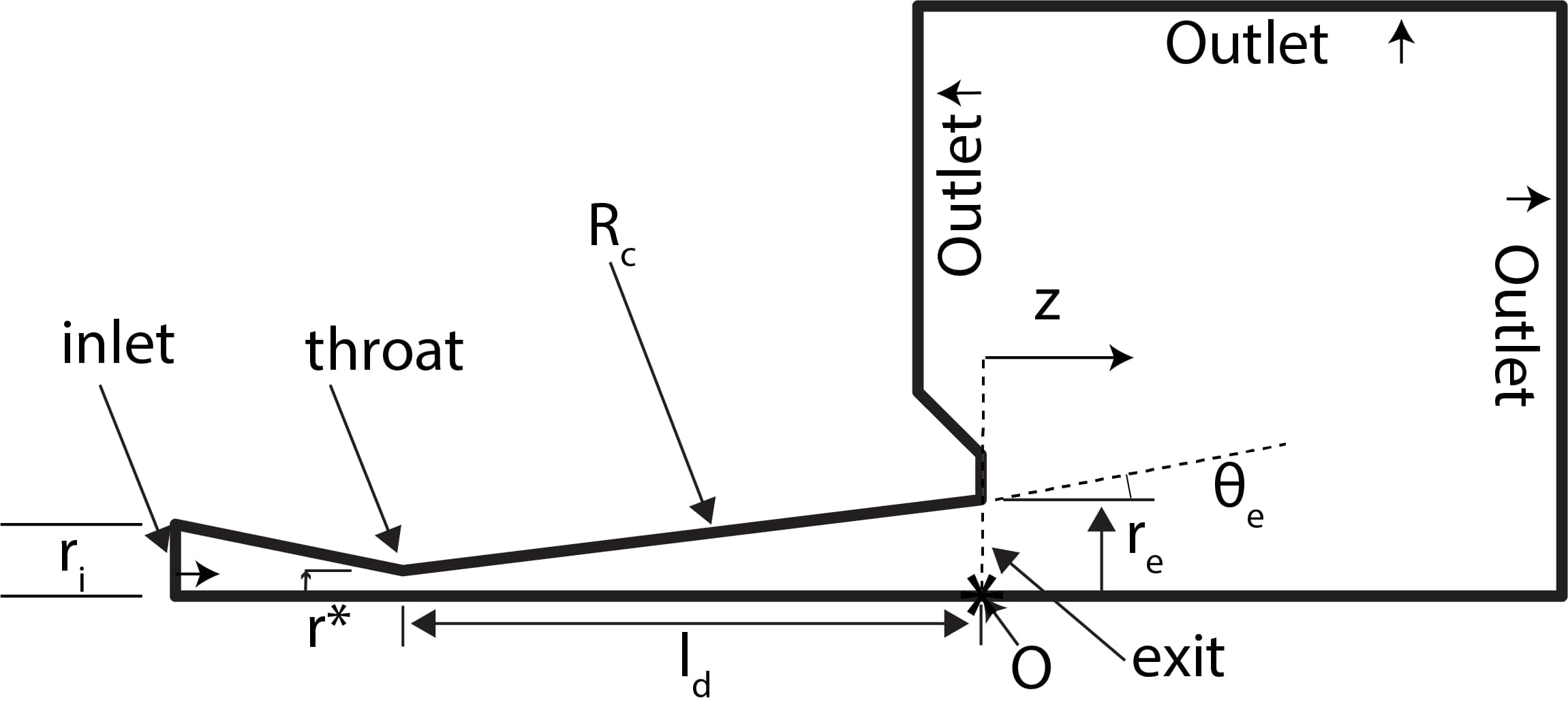}
    \caption{Axisymmetric domain used for simulations}
    \label{fig:2D_Domain}
\end{figure}

 This section covers the basic theory behind supersonic nozzle design, the simulated nozzle geometries and the simulation methods employed to model the gas jets produced by each nozzle. The variables shown in the axisymmetric domain in Fig. \ref{fig:2D_Domain} are referenced throughout the paper and defined as:
\begin{enumerate}
    \item $r^*, r_i, r_e$: nozzle throat, nozzle inlet and nozzle exit radius. Corresponding diameters defined the same way, ($d^*, d_i, d_e)$
    \item $l_d$: length of diverging section
    \item $z$: normal distance from nozzle exit. $z$ $<$ 0 means inside of nozzle
    \item $R_c$: radius of curvature of diverging section
    \item $\theta_e$: nozzle exit half-angle
    \item O: origin of coordinate system
    \item Outlet: where the gas exits in the domain
\end{enumerate}

\subsection{Nozzle Geometries and Design}
For each simulated nozzle geometry, 1D isentropic flow theory was first used to choose the desired $d_e$ and $d^*$ with various radii of curvature, $R_c$, and lengths, $l_d$, being chosen after. The 1D isentropic flow model approximates important flow parameters such as density $\rho$ and Mach Number $M$ and relates these parameters to the nozzle geometry through the cross section area $A$ \cite{CompFlow, gasdynamicsbook, Nasareport, Semushin}. Any variable with subscript 0 indicates a stagnation quantity, referring to the quantities of the gas in the gas bottle. Any variable with superscript * refers to a quantity at the nozzle throat. $\kappa$ is the specific heat ratio of the gas. The isentropic flow equations used to design the nozzles were:

\begin{equation}
    \frac{\rho}{\rho_0} = (1 + \frac{\kappa - 1}{2}M^2)^{-\frac{1}{\kappa - 1}}
    \label{eq: rhoisen}
\end{equation}

\begin{equation}
    \frac{A}{A^*} = \frac{1}{M}\left[\frac{2}{\kappa + 1}(1 + \frac{\kappa - 1}{2}M^2)\right]^{\frac{\kappa + 1}{2(\kappa - 1)}}
    \label{eq: areamach}
\end{equation}

 Eqn. (\ref{eq: areamach}) was used to calculate the exit Mach number for each nozzle, optimizing the nozzle geometry for a desired exit Mach number, $M_e$. All nozzles were chosen to have the same exit diameter, $d_e$ = $3$ mm, to match (including consideration of the gas flow dynamics to the interception point of the laser) the accelerating structure with laser parameters for dephasing, depletion and diffraction. The large $d_e$ also lessens the effect of boundary layers on the flow as opposed to sub-mm scale nozzles \cite{gasjetSchmid}. For $d^*$ = 0.6 mm, isentropic calculations yield $M_e$ = 7.1 whereas for $d^*$ = 0.8 mm, $M_e$ = 5.7. The higher exit Mach numbers were chosen as a sequel to previously lower $M_e$ nozzles used \cite{liona, kkswan, haien} as high exit Mach numbers correspond to more flat-top density profiles \cite{Semushin}. Both $d^*$ were also chosen so that, with a backing pressure of 500 psi, isentropic exit density $\rho_e$ would be on the order of $10^{19}$ $cm^{-3}$, the optimal LPA density range \cite{gasjetSchmid, cgrth}. $\kappa = 5/3$ for Helium (He) and Argon (Ar).

\begin{table*}[htbp!]
\caption{\label{tab: gases} Characteristic thermodynamic and flow values for He and Ar}
\begin{ruledtabular}
\begin{tabular}{cccccccc}
 Gas & $C_p$ [J/kg*K] & $k$ [W/m*K] & $A$ & $n$ & $\mu_0$ [Pa*s] & $S$ [K] & $T_0$ [K]\\ \hline
He & 5193 & 0.152 & 4.078$\times 10^{-7}$ & 0.6896\\
Ar & 520.64 &  0.0158 & &  & 2.125$\times 10^{-5}$ & 144.4 & 273.11\\
\end{tabular}
\end{ruledtabular}
\end{table*}

Three different curvatures were simulated, with their geometry parameters defined below:
\begin{enumerate}
    \item Concave "Bell" ($R_c>0$) nozzle: $d^*$ = $0.6$ mm, $R_c$ = $+31$ mm
    \item Straight conical nozzle: $d^*$ = $0.8$ mm, $R_c$ = $\infty$, $l_d$ varied
    \item Convex "Trumpet" ($R_c<0$) nozzle: $d^*$ = $0.8$ mm, $R_c$ varied, $l_d$ varied
\end{enumerate}

To compare the effect of curvature, three of the simulated nozzles, one from each curvature, were constrained to have $l_d$ = $9$ mm with all other parameters also kept constant. The bell nozzle having $d^*$ = 0.6 mm as opposed to $d^*$ = 0.8 mm was found to not have significant impact on the qualitative features observed, as shown in Fig. S3 of the supplementary document. The straight and trumpet nozzles were chosen to have greater $d^*$ to loosen manufacturing constraints. The inlet diameter $d_i$, which has little effect on the gas jet \cite{Atkinson}, was chosen to match the valve diameter of $2.24$ mm. The diverging section length, $l_d$, was varied to optimize the trumpet and straight nozzles. In past studies of the straight nozzle, $l_d$ was approximately optimized using the "$1/M_e$" condition, which matched exit half-angle, $\theta_e$, to the Mach angle, also known as the shock angle, of $\sin^{-1}{(1/M_e)}$, to minimize the shock intensity \cite{Semushin, MKrish, cgrth}. The radius of curvature of the bell nozzle, $R_c$ = $+31$ mm was chosen to demonstrate the effect of the bell. The trumpet $R_c$ was varied to find the optimal trumpet geometry for producing flat-top density profiles.

\subsection{Simulation Methods}
The gas jet simulations were performed using the computational fluid dynamics (CFD) program ANSYS Fluent, which provides a range of numerical solvers for the Navier-Stokes, continuity and energy equations \cite{fluent}. Both 2D-axisymmetric and 3D simulations were performed. While 2D-axisymmetric simulations are computationally less expensive and can be more refined, 3D simulations model turbulence and flow more accurately. Thus, density maps were extracted from the 3D simulations whereas 2D-axisymmetric simulation results were used to resolve finer features such as shocks. The domain used for the 2D-axisymmetric simulations is shown in Fig. \ref{fig:2D_Domain}. The mesh for the domain consisted of about $2.5 \times 10^5$ quadrilateral elements, also called cells. Most of the cells were close to being perfectly square, which is ideal for CFD simulations \cite{flutheo}. The solver settings were the exact same as the 3D simulations settings. The boundary conditions (BCs) were also the same except with an added "axis" BC due to the axisymmetric nature. The axisymmetric profile of the 3D domain had the same nozzle profile as the 2D domain but was smaller in outlet area by 75\% in order to allow for a more refined mesh with the cells being closer to cubes, which is ideal for CFD simulations \cite{flutheo}. The profile was revolved to create the 3D domain. The mesh for 3D simulations contained $4.5 \times 10^5$ cells, with the average skewness being 0.062. The average orthogonality is 0.983, and the average aspect ratio is 3.74. An implicit coupled density-based steady-state solver was used with double precision accuracy. Turbulence was modeled using the $k-\omega$ shear stress transport (SST) model, which models turbulence both near and far from the walls well \cite{turbsst}. Spatial discretization was done with the Least Squares Cell-Based (LSCB) method given its better accuracy, stability and speed compared to other provided methods such as the Green-Gauss Node Based (GGNB) method \cite{flutheo, grad}. Turbulence was modeled with a third order method while flow was modeled with a second order method to yield more accurate solutions. The gases tested were Helium (He) and Argon (Ar), modeled by the ideal gas equation of state. Heat capacity $C_p$ and thermal conductivity $k$ were assumed to be constant for both gases. Viscosity for He was modeled with the power law model, $\mu$ = $AT^n$, while for Ar, the Sutherland 3-coefficient model was employed, $\mu$ = $\mu_0 (\frac{T}{T_0})^{\frac{3}{2}} \frac{T + S}{T_0 + S}$. The power law was interpolated from past empirical data \cite{HeData}. Otherwise ANSYS Fluent's default parameters imported from the NIST database  were used \cite{flutheo}. The parameters for the gases are shown in table \ref{tab: gases}. The following BCs were applied:
\begin{enumerate}
    \item inlet: Pressure BC of 500 Psi. Temperature set at 300 K. 
    \item outlet: Pressure BC of 1 milliTorr, the ambient vacuum pressure, $P_{amb}$. Temperature set at 300 K. 
    \item wall: no-slip condition with no roughness assumed for simplicity. 
\end{enumerate}

These conditions were chosen to closely represent typical experimental conditions. From each nozzle simulation, the 2D density map along the diameter of the nozzle exit was extracted for the output gas jet plume where z $>$ 0. Density profiles at mm-scale distances from the nozzle exit were then obtained, as are used for typical LPA experiments.

\begin{figure*}[h!tpb]
    \centering
    \includegraphics[width = \textwidth]{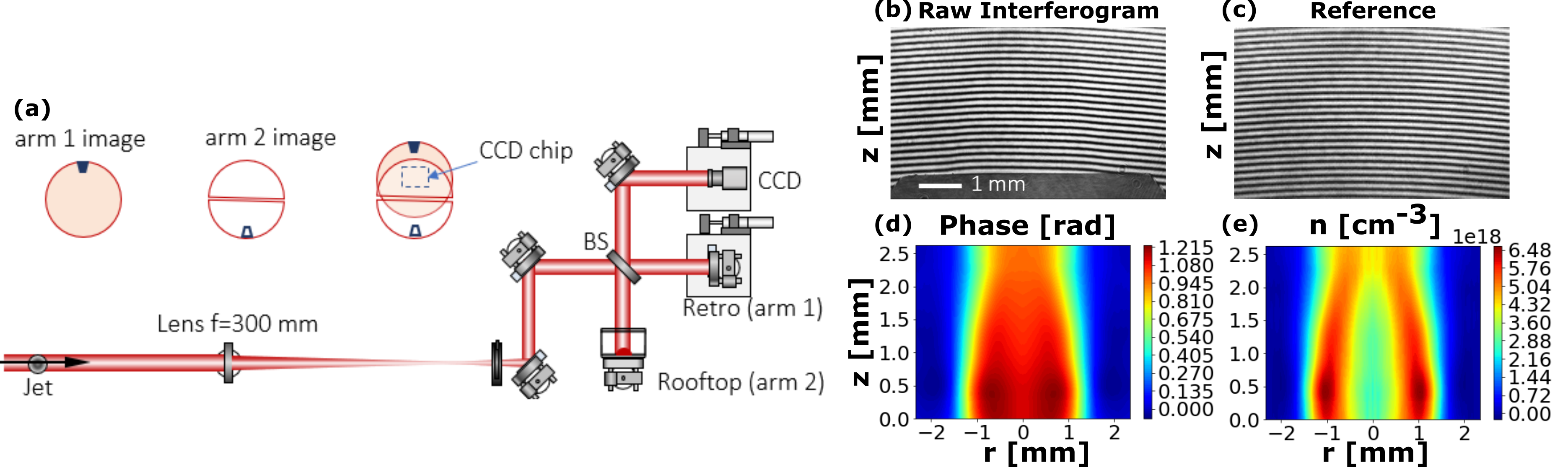}
    \caption{(Color) \textbf{(a)} Interferometry setup. The raw interferogram, in \textbf{(b)}, is compared with the reference interferogram, in \textbf{(c)}, to extract the corresponding phase distribution, shown in \textbf{(d)}. The phase distribution is then converted to a density map, $n$, by performing an Abel Inversion, shown in \textbf{(e)}.} 
    \label{fig: IFSetup}
\end{figure*}

\begin{table*}[htpb!]
\caption{\label{tab: isentropcomp} Comparison of average exit density $\rho_e$, exit Mach number $M_e$, and throat Mach number $M^*$ between simulation results and 1D isentropic flow predictions for all three nozzle geometries}
\begin{ruledtabular}
\begin{tabular}{ccccccc}
Nozzle & Sim. $\rho_e$ [$cm^{-3}$] & 1D $\rho_e$ [$cm^{-3}$] & Sim. $M^*$ & 1D $M^*$ & Sim. $M_e$ & 1D $M_e$\\ \hline
Bell & 1.09 $\times 10^{19}$ & 1.11 $\times 10^{19}$ & 0.83 & 1 & 6.73 & 7.09\\
Straight & 2.27 $\times 10^{19}$ & 2.00 $\times 10^{19}$ & 0.80 & 1 & 5.26 & 5.74\\
Trumpet & 2.37 $\times 10^{19}$ & 2.00 $\times 10^{19}$ & 0.86 & 1 & 5.09 & 5.74\\
\end{tabular}
\end{ruledtabular}
\end{table*}

\section{\label{sec: Experiment Method} Experimental Method}
Three nozzles, one from each curvature, out of all the simulated nozzle designs were manufactured and experimentally characterized using neutral density interferometry. Nozzle costs were not significantly different from previously manufactured nozzles. All three manufactured nozzles were made of aluminum and had $l_d$ = 9 mm and $d_e$ = 3 mm. The bell nozzle had $d^*$ = 0.6 mm and $R_c$ = +31 mm while the trumpet nozzle had $d^*$ = 0.8 mm, like the straight nozzle, and $R_c$ = -100 mm. A Michelson interferometer was used to characterize the gas jet density field of each nozzle. The setup is shown in Fig. \ref{fig: IFSetup}(a). The experiment was performed using the Hundred Terawatt Thomson (HTT) laser system at the Berkeley Lab Laser Accelerator (BELLA) center, specifically using the 1 Hz mJ-level probe laser beam, which is split after the first main amplifier stage and independently compressed to 40 fs with 800 nm center wavelength and 15 mm beam diameter. It propagates through the gas jet, imaging the gas jet plane, and is focused by a f/\# = 20 lens. The beam is then split by a beamsplitter (BS) which directs the reflected beam into a retroreflecting roof mirror pair (image) and the transmitted beam into a 0-degree high reflective mirror (reference). Both beams are then recombined by the same BS and imaged onto a CCD camera calibrated to 2.64 $\mu$m/pixel in the gas jet plane with a field of view of 4.6 x 3 mm$^2$ and a resolution of 15 $\mu$m. The gas jet nozzle is mounted onto a solenoid valve that allows for continuous or pulsed gas delivery. Shot-to-shot fluctuation was observed to be low at $\sim$ 2\%, with $1/3$ of this being from the imaging system and laser pulse fluctuations (determined by analyzing the variations of reference scans that contained no gas flow). Ar gas was used due to its higher index of refraction compared to He. 

The laser beam passing through the gas flow experiences a phase shift, quantified by fringe shifts on the resulting interferogram compared to the reference. The phase shift distribution is reconstructed from the fringe shifts. Because the nozzles are axisymmetric, an Abel inversion is performed to radially symmetrize the phase shift distribution and extract the the index of refraction distribution, $n$. The Abel inversion was performed by applying the mathematical algorithm delineated in previous papers \cite{Vmalka, gasjetSchmid, BockastenAbel} through MATLAB. The phase shift is related to $n$ through the following equation, 

\begin{equation}
    \Delta\phi(r) = k\int (n(r, l) - 1)dl = \frac{2\pi}{\lambda}\int (n(r, l) - 1)dl,
    \label{eq: phaseshifteqn}
\end{equation}

where $\Delta \phi$ is the phase shift, $k$ is the wavenumber and $\lambda$ is the laser wavelength \cite{hutch}. The variation of the index of refraction, $n$, is then related to gas atom or molecule number density, $N$, through the Lorentz-Lorenz equation, 
\begin{equation}
    N = \frac{3}{4\pi\alpha}\frac{n^2 - 1}{n^2 + 2},
    \label{eq: LLeqn}
\end{equation}
where $\alpha$ is the mean polarizability, defined as $\alpha = \frac{3A}{4\pi N_A}$ \cite{earlystudyJLHen, born}. $A$ is the molar refractivity and $N_A$ is Avogadro's number. Substituting the relation for $\alpha$, we get:
\begin{equation}
    N = \frac{N_A}{A}\frac{n^2 - 1}{n^2 + 2}
    \label{eq: LLsimp}
\end{equation}
Index of refraction data for Ar was used to calculated the molar refractivity, A, of Ar using Eqn. (\ref{eq: LLsimp}), found to be (4.138 $\pm$ 0.012) $\times$ $10^{-6}$ m$^3$/mol \cite{Armolar}.

To maximize interferometry signal, scans for each nozzle were taken at the maximum regulator pressure of 1000 psi. The interferograms were averaged and converted to 2D density maps, outlined in Fig. \ref{fig: IFSetup}(b)-(e). Inherent noise close to the axis from the Abel inversion algorithm \cite{earlystudyJLHen} coupled with uncertainty of the nozzle axis led to larger apparent density fluctuations closer to the nozzle axis. The density maps and density lineouts extracted from the maps were compared to simulation results.

\begin{figure*}[h!tpb]
    \includegraphics[width = \textwidth]{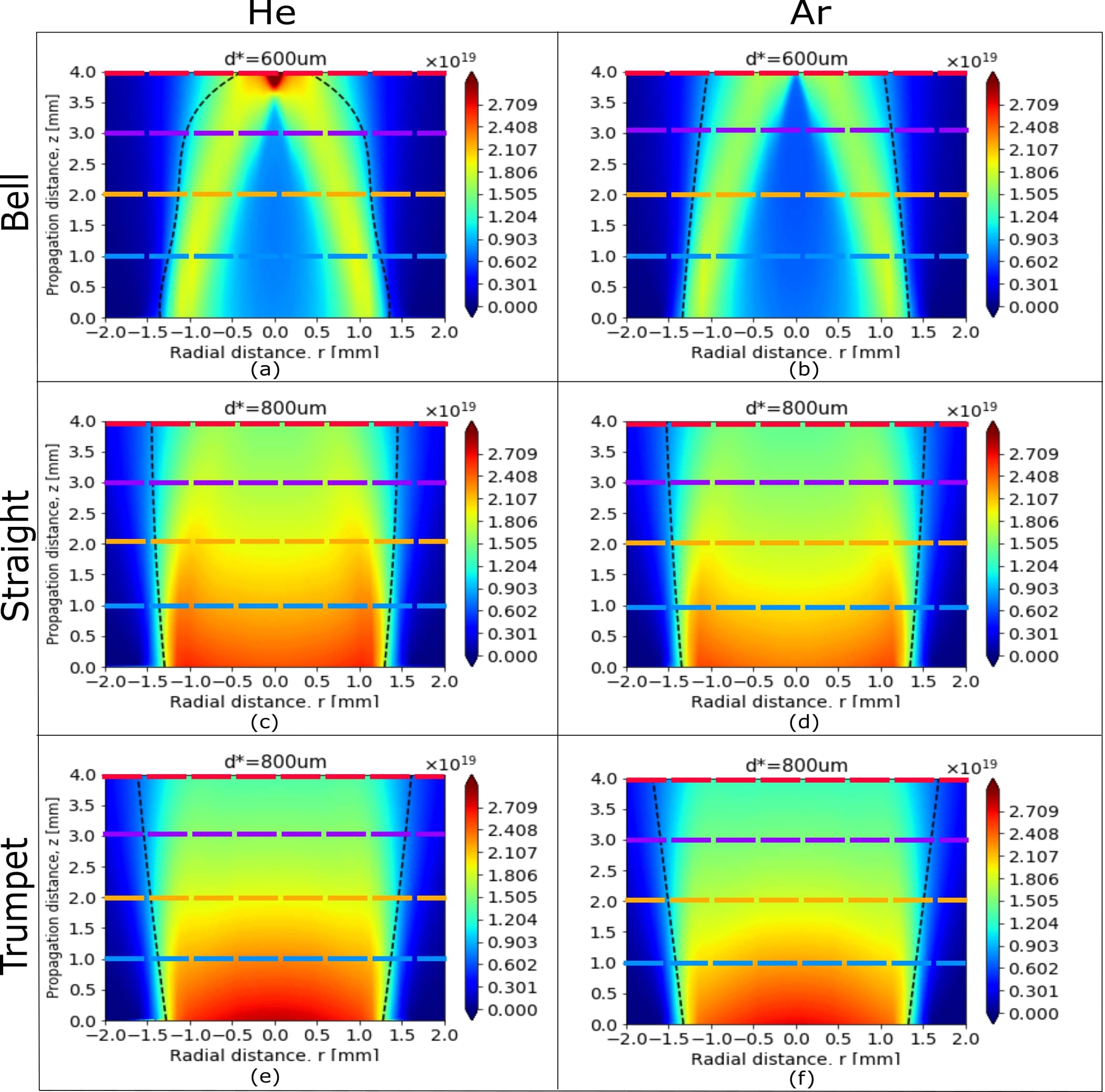}
    \caption{(Color) Atomic density maps extracted from 3D simulations for the three nozzle geometries. Density in units of cm$^{-3}$. The black dashed lines on each map denote the FWHM of the density profiles. Note the focusing effect of the bell nozzle, where the FWHM decreases noticeably close to the shock diamond. Density profiles at z = 1, 2, 3 and 4 mm are extracted from each map, along the colored dashed lines drawn. The left column shows the extracted maps when using He as the gas: \textbf{(a)} 600 $\mu$m Throat Bell Nozzle \textbf{(c)} 800 $\mu$m Throat Straight Nozzle \textbf{(e)} 800 $\mu$m Throat Trumpet Nozzle. The right column shows the maps for Ar as the gas:
    \textbf{(b)} 600 $\mu$m Throat Bell Nozzle \textbf{(d)} 800 $\mu$m Throat Straight Nozzle \textbf{(f)} 800 $\mu$m Throat Trumpet Nozzle } 
    \label{fig: DensMaps}
\end{figure*}

\begin{figure*}[h!tpb]
    \includegraphics[width = \textwidth]{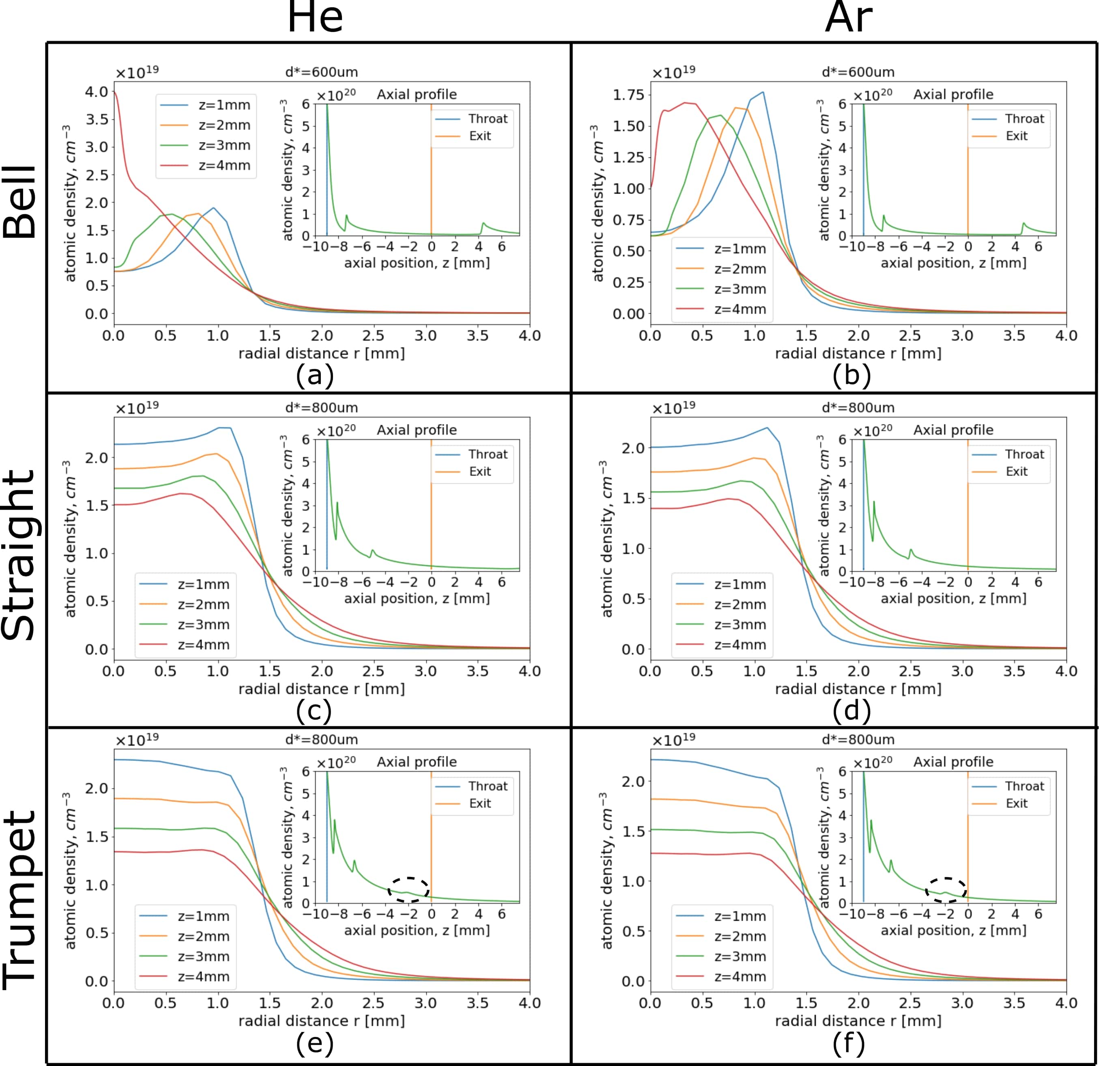}
    \caption{(Color) Axisymmetric density lineouts extracted from 3D simulations for the three nozzle geometries. The inset plots show the density lineouts along the nozzle axis extracted from 2D simulations, with the sharps discontinuities indicating presence of shock diamonds. The left column shows the extracted profiles when using He as the gas: \textbf{(a)} 600 $\mu$m Throat Bell Nozzle \textbf{(c)} 800 $\mu$m Throat Straight Nozzle \textbf{(e)} 800 $\mu$m Throat Trumpet Nozzle. The right column shows the profiles for Ar as the gas:
    \textbf{(b)} 600 $\mu$m Throat Bell Nozzle \textbf{(d)} 800 $\mu$m Throat Straight Nozzle \textbf{(f)} 800 $\mu$m Throat Trumpet Nozzle} 
    \label{fig: DensLineouts}
\end{figure*}

\section{\label{sec: Results} Results}

\subsection{\label{sec: compcurv} Effect of Diverging Section Curvature}

\begin{table*}[htpb!]
\caption{\label{tab: FWHM} FWHM of the simulation density profiles for all three geometries across both gases. Units in mm.}
\begin{ruledtabular}
\begin{tabular}{ccccccc}
 z [mm] &  He, Bell & He, Straight & He, Trumpet & Ar, Bell & Ar, Straight & Ar, Trumpet\\ \hline
2 & 2.27 & 2.81 & 2.92 & 2.43 & 2.91 & 3.01\\
3 & 2.09 & 2.87 & 3.08 & 2.31 & 2.99 & 3.19\\
4 & 0.75 & 2.89 & 3.23 & 2.05 & 3.05 & 3.37\\
\end{tabular}
\end{ruledtabular}
\end{table*}

2D density maps and radial density lineouts were extracted from each simulation for the manufactured geometries. The qualitative features between the maps and lineouts produced by each curvature were then compared. Test simulations were first done to confirm that $d^*$ being altered between $0.6$ and $0.8$ mm had little effect on simulation results. The simulated 2D density maps for the three manufactured geometries are shown in Fig. \ref{fig: DensMaps} while corresponding simulated density lineouts are shown in Fig. \ref{fig: DensLineouts}. The similarity in gas jet behavior and shock features between He and Ar, with the density maps and profiles for all three geometries having like shapes between the two gases, matches expectations in accordance with the isentropic flow equations, Eqns. (\ref{eq: rhoisen}) and (\ref{eq: areamach}), as He and Ar have the same specific heat ratio, $\kappa$ = $5/3$. A further comparison, shown in Table \ref{tab: isentropcomp}, between 1D isentropic flow predictions, calculated from using Eqn. \ref{eq: rhoisen} and \ref{eq: areamach}, and simulation results, indicates that the simulated gas flows roughly follow 1D isentropic flow as expected. The differences are due to the 1D model not accounting for shock features and losses from 2D and 3D effects \cite{Semushin}. 

\begin{table*}[htpb!]
\caption{\label{tab: flatnesstable} Profile flatness and sharpness comparison between the straight and trumpet nozzles across both gases. $l_{flat}$ is the flat-top length. $\delta_l$ is the profile gradient length. $(\sigma_n)/\langle n \rangle$ is the density standard deviation divided by the mean density in the plateau region.}
\begin{ruledtabular}
\begin{tabular}{c|ccc|ccc|ccc}
\hline
&\multicolumn{3}{c|}{z = 2 mm}&\multicolumn{3}{c|}{z = 3 mm}&\multicolumn{3}{c}{z = 4 mm}\\
Gas, Nozzle & $l_{flat}$ [mm] & $\Delta l$ [mm] & $(\sigma_n)/\langle n \rangle$ [\%] & $l_{flat}$ [mm] & $\Delta l$ [mm] & $(\sigma_n)/\langle n \rangle$ [\%] & $l_{flat}$ [mm] & $\Delta l$ [mm] & $(\sigma_n)/\langle n \rangle$ [\%] \\ \hline
He, Trumpet & 2.39 & 0.70 & 1.80 & 2.37 & 0.96 & 1.86 & 2.28 & 1.27 & 1.83\\ 
He, Straight & 2.31 & 0.64 & 2.78 & 2.15 & 0.94 & 2.83 & 1.93 & 1.25 & 2.87\\ 
Ar, Trumpet & 2.45 & 0.71 & 2.23 & 2.45 & 0.97 & 1.70 & 2.43 & 1.26 & 1.75\\ 
Ar, Straight & 2.45 & 0.63 & 2.67 & 2.27 & 0.93 & 2.51 & 2.09 & 1.24 & 2.60\\
\end{tabular}
\end{ruledtabular}
\end{table*}

Observed from the density maps in Figs. 4(a), (b), the bell geometry creates a focusing effect that places a shock diamond at around $z_d$ = 3.7 mm for He and $z_d$ = 4 mm for Ar, right in the region of interest. This causes large density fluctuations, where the density is much lower at positions before the shock diamond, $z < z_d$, compared to points closer to the to the shock diamond position, $z \approx z_d$, preventing effective formation of flat-top density profiles, seen in Figs. 5(a), (b). For example, for the case of He in Fig. \ref{fig: DensLineouts}a, the density profiles for z = 1, 2 and 3 mm all have an "M" shape, dipping down to a density of $\sim 8 \times 10^{18}$ cm$^{-3}$. These "M" shape profiles have also been observed in past studies on the bell nozzle \cite{MKrish}, yielding uneven profiles \cite{Semushin}. The z = 4 mm profile, closer to the shock diamond, displays a density spike up to $\sim 4 \times 10^{19}$ cm$^{-3}$, about 5 times the density dip before the shock diamond. This density spike produced by the bell can be useful for creating short, high-density gas targets. This focusing effect, also seen in past studies \cite{MKrish, bell}, is observed in the Full Width Half Maximum (FWHM) of the profiles, listed in Table \ref{tab: FWHM}, where the FWHM decreases significantly for the z = 4 mm profile of the bell nozzle. The large fluctuations and focusing effect are caused by the formation of standing shock waves from the nozzle throat to its outer region due to its exit pressure and ambient pressure not matching \cite{liona, MusinskiGasJet}. Transitioning to the straight nozzle map, shown in Fig. \ref{fig: DensMaps}(c), (d), the shock diamond is no longer intensely concentrated to a point, demonstrating a weaker focusing effect as observed before \cite{MKrish}. While the straight nozzle profiles, shown in Figs. 5(c), (d), are closer to the flat-top shape, noticeable density variations along potential laser interaction paths remain, matching past observed density lineouts of straight nozzles \cite{gasjetSchmid, froulaIF}. When the curvature is inverted to the trumpet geometry, shown in Figs. 4(e), (f), the shock diamond is suppressed as the trumpet geometry reverses the focusing effect of the bell curvature. The shock suppression of the trumpet nozzle prevents large density fluctuations at the output, removing the fluctuations at the plateau edges and leading to flatter plateau regions, seen in Figs. 5(e), (f). 

For LPA applications, the plateau length, plateau uniformity and profile sharpness are all important characteristics of the density profile. The plateau length sets the length scale of the electron acceleration. A uniform plateau is essential for LPA since small density perturbations in the plateau can cause electron dephasing and ruin acceleration \cite{cgrth} as well as lead to unwanted self-injection during acceleration \cite{SBulanov, selfinjPhysLett}. Tighter density fluctuation tolerances become even more important as higher quality beams are required from LPA. On the other hand, sharper profiles, meaning smaller density gradients, are also desirable as they lessen the deceleration length scale of the electrons \cite{Semushin}. To compare all three characteristics, the flat-top length, normalized density standard deviation within the plateau region and density gradient length between the straight and trumpet nozzles are shown in Tab. \ref{tab: flatnesstable}. The flat-top length, $l_{flat}$, is defined as the length over which the density lies within 10\% of the maximum density \cite{cgrth}. The profile gradient length, $\Delta l$, is the length over which the density rises from 10\% to 90\% of the density at r = 0 \cite{gasjetSchmid}. While both nozzles give similar flat-top lengths, the straight nozzle produces sharper profiles, with its profile gradients being smaller. On the other hand, the trumpet nozzle yields flat-top regions with less deviations from the mean plateau density. Thus, the trumpet nozzle creates a flatter profile at the cost of wider profile gradients, compared to the straight nozzle. Therefore, the trumpet nozzle can be applied for situations where density fluctuation tolerances are tighter whereas the straight nozzle can be used for when density gradients are required to be sharper. The simulation density gradient thicknesses agree well with the 1D isentropic theory estimates, where $\Delta l$ = $2z/M_e$ \cite{cgrth, gasdynamicsbook}, with the estimates being $\Delta l$ $\approx$ 0.70, 1.05, 1.40 mm for z = 2, 3 and 4 mm respectively.

The different shock features between the curvatures can also be observed from the density lineouts along the nozzle axis, shown in the inset plots of Fig. \ref{fig: DensLineouts}. Sudden spikes in the axial profiles correspond to shock diamonds created from standing shock waves \cite{MusinskiGasJet}, which are formed due to a sufficiently high $P_{exit}/P_{amb}$, causing compression waves to coalesce into focused shocks \cite{liona}. This is typical for under-expanded jets, where $P_{exit} > P_{amb}$, and has been extensively studied \cite{underexp}. The position of the nozzle throat and exit are marked in the inset plots to indicate the relative positions of the shock diamonds. For the bell nozzle, the second shock diamond is $\sim$4 mm from the exit, outside of the nozzle and is comparable to the first shock in magnitude. When observing the straight nozzle's axial profile, this second shock diamond is pushed back into the region inside the nozzle between the throat and exit with no observable density spike in the region outside the nozzle, indicating a weaker focusing effect. For the trumpet nozzle, in addition to the second shock diamond being pushed back behind the exit, a third shock diamond, weaker in magnitude, is also pushed to sit behind the exit, exhibiting the trumpet nozzle's shock suppression. This third shock diamond is circled in the inset plots of Figs. 5(e), (f). 

\begin{figure*}[h!tpb]
    \includegraphics[width =\textwidth]{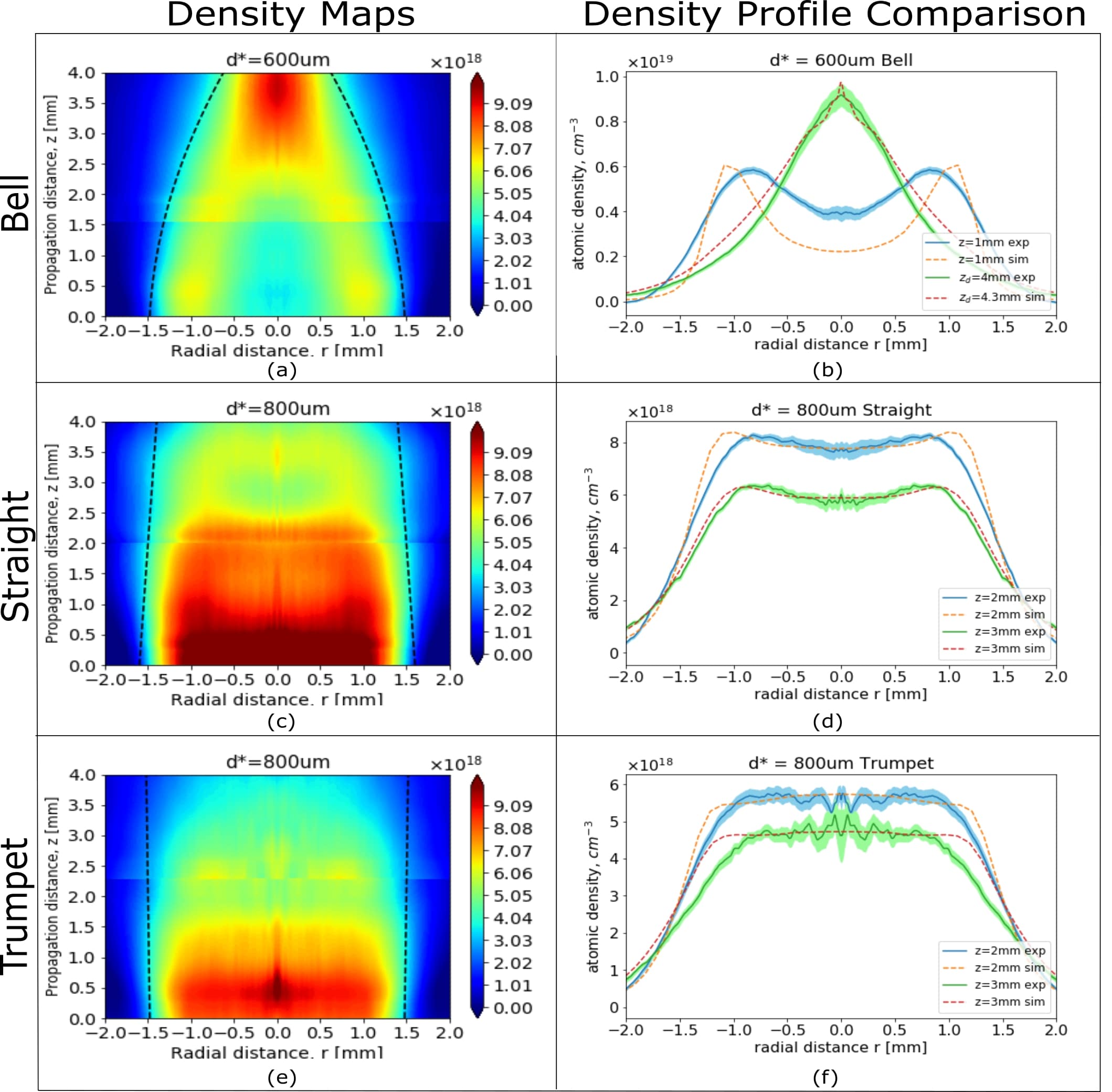}
    \caption{(Color) Gas used was Ar. Measured gas jet density maps for \textbf{(a)} bell nozzle, \textbf{(c)} straight nozzle, \textbf{(e)} trumpet nozzle. The black dashed lines on each map denote the FWHM of the density profiles. Measured profiles are compared with simulation for the \textbf{(b)} bell nozzle, \textbf{(d)} straight nozzle, \textbf{(f)} trumpet nozzle. The simulation profiles are normalized to the measured profiles to compare shape. The shaded regions represent the RMS density fluctuations of each profile. The $z$ = 4 mm measured bell profile is compared with the $z$ = 4.3 mm simulation bell profile to compare the shape of the density spikes at the shock diamond.} 
    \label{fig: ExpMes}
\end{figure*}

These results can be scaled to other density regimes, such as densities with orders of magnitude of a few $10^{17}$ or $10^{20}$ $cm^{-3}$, by simply adjusting the backing pressure, $P_0$. Changing $P_0$ only alters the peak density, with the peak density scaling approximately linearly with $P_0$, and does not significantly affect the profile shape, as confirmed by results shown in Fig. S2 of the supplementary document as well as previous studies \cite{Semushin, Vmalka, haien, cgrth}. However, with different nozzle geometry parameters such as $d^*$ and $d_e$, both the bell \cite{Lemos} and straight nozzles \cite{gasjetSchmid, MKrish, Semushin} can produce uniform flat-top profiles with sharp edges. For example, increasing the exit diameter of a bell nozzle will push the shock diamond farther out from the nozzle exit, leading to flatter density profiles close to the nozzle exit as observed before for a bell nozzle with $d_e$ = 8 mm \cite{Lemos}. The straight nozzle can also be optimized to produce flat-top profiles using the Mach angle condition as a lower limit for the nozzle's exit half-angle, for sub-mm scale nozzles \cite{gasjetSchmid} and mm-scale nozzles \cite{MKrish, Semushin}. However, the effects of nozzle curvature on shock features observed here still hold for axisymmetric mm-scale nozzles as also shown in past studies \cite{MKrish, Semushin, Lemos, gasjetSchmid, froulaIF}. 

\subsection{\label{sec: compres} Comparison of Simulation Results with Experimental Measurements}

The experimental measurement results are shown in Fig. \ref{fig: ExpMes}. For each nozzle, the interferograms taken closer and farther from the nozzle exit were concatenated to yield the full density map. The measured density maps matched well with simulation maps in shape and shock features, such as the FWHM lines. The strong focusing effect in the measured density field of the bell nozzle, shown in Fig. \ref{fig: ExpMes}(a), is observed, where the density before the shock diamond is low but spikes up as closer to the shock diamond. The shock suppression of the trumpet nozzle is similarly observed, shown in Fig. \ref{fig: ExpMes}(c) and (e) respectively.

The actual pressure delivered to the nozzle inlet is unknown and likely lower than set by the gas regulator due to the lossy connections between the valve and regulator. This explains why the measured density is lower than that from simulation, which treats the inlet pressure as the same as the regulator pressure. We confirmed backing pressure only changes the peak density and not the normalized profile shape as shown in Fig. S2 of the supplementary document, agreeing with previous studies \cite{Semushin, Vmalka, cgrth, haien}. Therefore, the simulation profiles were normalized to the measured profiles for profile comparison. The measured profiles demonstrated the qualitative features predicted by simulation. For the bell nozzle profiles in Fig. \ref{fig: ExpMes}(b), the z = 2 mm measured profile shows a dip similar to the simulation profile. At z = 4 mm, the measured profile displays a spike, characteristic of the large spikes observed in simulation when close to the shock diamond. The measured straight nozzle profiles contained the slight density dips observed in simulation, preventing them from being flat-top. The measured trumpet nozzle profiles at z = 2 and 3 mm were flat-top as predicted by simulation. For the trumpet and straight nozzles, the simulation profiles are broader at the edges compared to the experimental profiles, which can be explained by the inability of our simulations to accurately model the complex fluid dynamics near the wall. ANSYS Fluent is limited in its assignment of the wall boundary condition, with the user being able to set a constant wall roughness and set either a no-slip (zero velocity at wall) condition or a constant shear stress condition \cite{flutheo}. Manufacturing effects often cause various tool marks that lead to inconsistent roughness values and the varying surface roughness, which can range from tens to hundreds of microns, impacts the resulting density profiles \cite{cgrth}. Our simulations set wall roughness to 0 for simplicity as we were unable to model the roughness patterns of the nozzles, which leads to a growing boundary layer developing along the wall. However, inconsistent roughness of the wall surface can cause particles to lose adhesion (known as wetting) to the surface at certain regions, leading to wall slip, where the fluid velocity is not zero at the wall \cite{wallslip}. Thus, the wall roughness and complex near-wall fluid dynamics can alter the boundary layer formation and in turn influence the density gradient thicknesses and profile edges \cite{gasjetSchmid}. Furthermore, the fluid behavior near the wall can be complex and would require molecular dynamics simulations to more accurately model \cite{knudsenwallslip}, especially since the wall is curved \cite{curvedwallslip}. These factors explain the observed profile edge discrepancy between the measured and simulated density profiles. All profiles had relatively small RMS fluctuations, indicated by the shaded regions around the measured profile in Fig. \ref{fig: ExpMes}(b), (d) and (e). The RMS fluctuations were determined by taking the standard deviation in density of each data point in the profile across hundreds of shots.

\subsection{Optimization of the Trumpet Geometry \label{sec: opttrump}}

\begin{figure}[H]
    \centering
    \includegraphics[width = 0.5\textwidth]{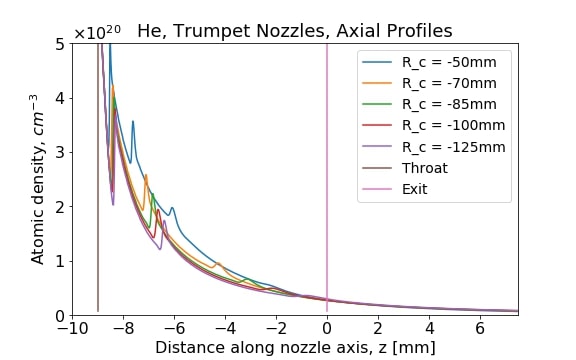}
    \caption{(Color) Density profiles along the nozzle axis for trumpet geometries with various $R_c$. Gas used was He. Nozzle throat and exit are marked.}
    \label{fig: tuningRc}
\end{figure}

Multiple trumpet geometries were simulated to optimize the flat-top profiles produced by the nozzle. The optimization procedure was then compared to the straight nozzle optimization process. The trumpet geometry's shock suppression does not automatically guarantee flat-top density profiles. The strength of the trumpet nozzle's shock suppression is inversely related to $R_c$'s magnitude, shown in Fig. \ref{fig: tuningRc}. For example, for the case of $R_c$ = -50 mm, the third shock diamond is much closer to the throat than for trumpet geometries with larger $R_c$. On the other hand, for $R_c$ = -125 mm, the third shock diamond is closer to the exit than for the other radii of curvatures. A larger curvature, meaning a smaller $|R_c|$, leads to shock diamonds being pushed further back into the diverging section, yielding a stronger suppression.

\begin{figure}[H]
    \centering
    \includegraphics[width = 0.5\textwidth]{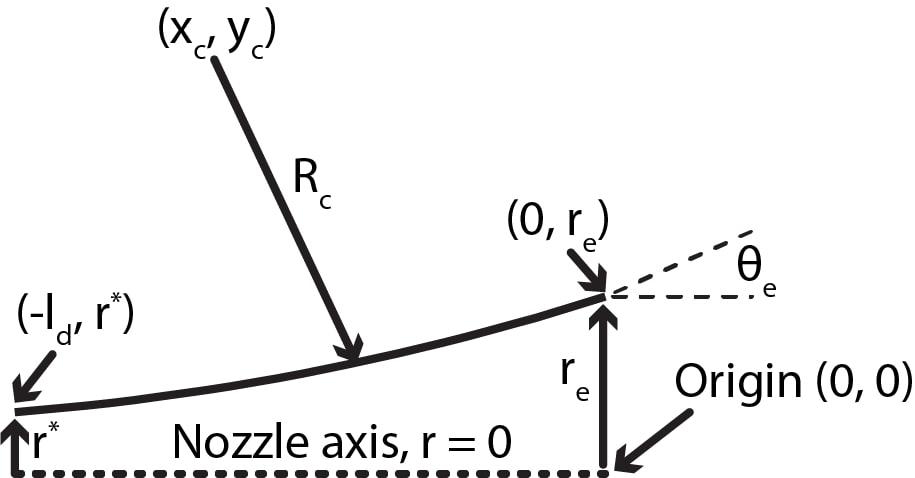}
    \caption{The trumpet nozzle's diverging section. Important parameters labeled.}
    \label{fig: halfanglecond}
\end{figure}

Because $d^*$ and $d_e$ are constrained, the optimal trumpet geometry involved finding the right combination of $R_c$, $l_d$ and $\theta_e$. The following two equations can be written to define the two endpoints of the arc, corresponding to Fig. \ref{fig: halfanglecond}:

\begin{equation}
    x_c^2 + (r_e - y_c)^2 = R_c^2; \quad (x_c + l_d)^2 + (r^* - y_c)^2 = R_c^2
    \label{eq: arceqn}
\end{equation}

The slope of the nozzle exit is then $\frac{dy}{dx}$ = $\frac{-(l_d - x_c)}{\Delta r - y_c}$. We define the variable $\Delta r = r_e - r^*$ to simplify the expression. Solving the two equations shown above, we find the exit slope to be:
\begin{equation}
    \scalebox{1}{
    $\frac{dy}{dx} = \frac{-(\Delta r)(\sqrt{-(\Delta r)^2((\Delta r)^2 + l_d^2)((\Delta r)^2 + l_d^2) - 4R_c^2} + (\Delta r)^2l_d + l_d^3)}{(\Delta r)^4 + (\Delta r)^2l_d^2 - l_d\sqrt{-(\Delta r)^2((\Delta r)^2 + l_d^2)((\Delta r)^2 + l_d^2) - 4R_c^2}}$}
    \label{eq: exitslope}
\end{equation}

This exit slope corresponds to a exit half-angle of $\theta_e$  = $\tan^{-1}{(\frac{dy}{dx})}$. The final optimization condition is then:
\begin{equation}
    \theta_e = \sin^{-1}{(1/M_e)} = \tan^{-1}{(\frac{dy}{dx}(d_e, d^*, l_d, R_c))}
    \label{eq: exitangle}
\end{equation}

This condition can be met by tuning the radius of curvature while keeping all other parameters the same. The calculated half-angles, $\theta_e$, corresponding to different radii of curvatures for the trumpet geometry used, where $M_e = 5.7$ and $l_d$ = 9 mm, are tabulated in Table \ref{tab: table2}. The optimal $R_c$ = -85 mm is labeled.

\begin{table}[H]
\caption{\label{tab: table2} Corresponding exit half-angles, $\theta_e$, for various $R_c$ for the  trumpet nozzle with $l_d$ = 9 mm and all other parameters kept constant. The optimal half angle, matching the Mach angle, is marked with an asterisk.}  
\begin{ruledtabular}
\begin{tabular}{lcdr}
\textrm{$R_c$ [mm]} & \textrm{$\theta_e$ [$^\circ$]}
\\
\colrule
-50 & 12.17\\
-70 & 10.68\\
-85 & 10.03*\\
-100 & 9.56\\
-125 & 9.05\\
\end{tabular}
\end{ruledtabular}
\end{table}

\begin{figure}[H]
    \centering
    \begin{subfigure}{0.5\textwidth}
    \centering
        \includegraphics[width = \textwidth]{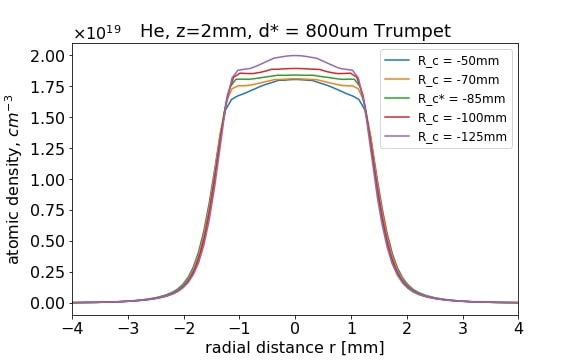}
        \caption{}
        \label{fig: z=2mmCurves}
    \end{subfigure}
    \begin{subfigure}{0.5\textwidth}
    \centering
        \includegraphics[width = \textwidth]{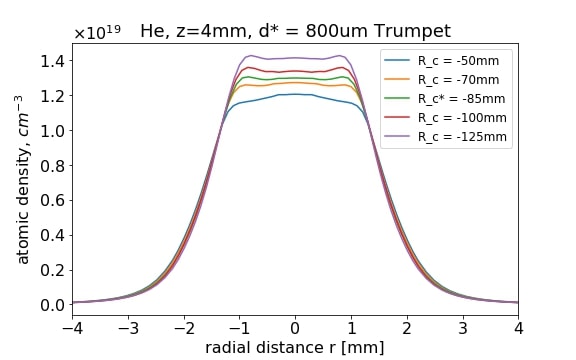}
        \caption{}
        \label{fig: z=4mmCurves}
    \end{subfigure}
    \caption{(Color) Simulation density profiles for trumpet geometry nozzles with various radii of curvature, $R_c$, at \textbf{(a)} z = 2 mm, and \textbf{(b)} z = 4 mm. All other parameters were kept the same and all nozzles had $M_e = 5.7$. Note the consistent flat-top profiles for nozzles that have exit half-angles close to the Mach angle. The optimal radius of curvature, $R_c$ = -85 mm, is marked with an asterisk.}
    \label{fig: tuningtrumpet}
\end{figure}

The radial density profiles for the various simulated $R_c$ in the trumpet geometry are shown in Fig. \ref{fig: tuningtrumpet}. For trumpet geometries with $\theta_e$ closer to the Mach angle, the density profiles remain flat-top, with z = 2 and 4 mm being shown as examples. The best consistency of the flat-top profiles is achieved at the optimal $R_c$ = -85 mm. With $R_c$ = -125 mm, the shock suppression is too weak, leading to more noticeable density variations. On the other hand, in the case of $R_c$ = -50 mm, the larger half-angle causes the output plume to diverge and disperse more, leading to a lower overall density and a nonuniform density profile. This optimization condition minimizes shocks by matching the $\theta_e$ to the Mach angle, analogous to the straight nozzle's Mach angle condition \cite{cgrth, Semushin, MKrish}.

\begin{figure}[H]
    \centering
    \includegraphics[width = 0.5\textwidth]{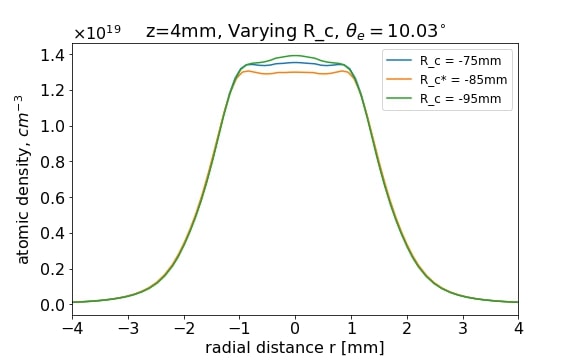}
    \caption{(Color) Simulation density profiles of the trumpet nozzles with various $R_c$ and thus $l_d$ with $\theta_e$ being held constant. The optimal radius of curvature, $R_c$ = -85 mm, is marked with an asterisk.}
    \label{fig: samethetadiffRc}
\end{figure}

Holding $l_d$ constant while $R_c$ is varied also changes the $\theta_e$, which can create interference between the respective effects of the two parameters. To isolate the effect of $R_c$, $\theta_e$ was held constant at the Mach angle $\theta_e$ = $10.03^{\circ}$ while $R_c$ was varied, which in turn changed $l_d$. As seen in Fig. \ref{fig: samethetadiffRc}, a difference in profile shape is observed between the three different $R_c$ geometries. In particular, the $R_c$ = $-95$ mm profile has a small bump, which can be explained by the weaker shock suppression. This indicates that while optimizing the trumpet nozzle's $\theta_e$ to be the Mach angle will approximately create flat-top profiles, further adjustment of $R_c$ is needed afterwards to ensure such profiles are created. 

\begin{figure}[H]
    \centering
    \includegraphics[width = 0.5\textwidth]{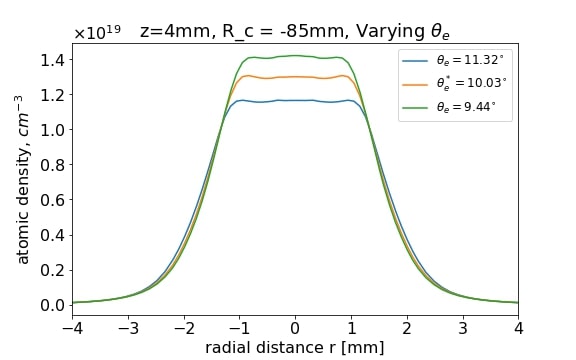}
    \caption{(Color) Simulation density profiles of the trumpet nozzles with various $\theta_e$ with $R_c$ held at the optimal -85 mm. Note how the flat-top profile shape is present even with different $\theta_e$. The Mach angle, $\theta_e$ = $10.03^{\circ}$, is marked with an asterisk.}
    \label{fig: sameRcdifftheta}
\end{figure}

On the other hand, for the trumpet nozzle, $\theta_e$ can be varied while $R_c$ is maintained at the optimal value of $-85$ mm. As observed in Fig. \ref{fig: sameRcdifftheta}, the flat-top shape is maintained at the optimal $R_c$ even though the $\theta_e$ varies. This further suggests that adjusting $R_c$ after optimizing $\theta_e$ to the Mach angle for the trumpet nozzle is more effective in producing consistent flat-top profiles than only optimizing the $\theta_e$.

\begin{figure}[H]
    \centering
    \begin{subfigure}{0.5\textwidth}
    \centering
        \includegraphics[width = \textwidth]{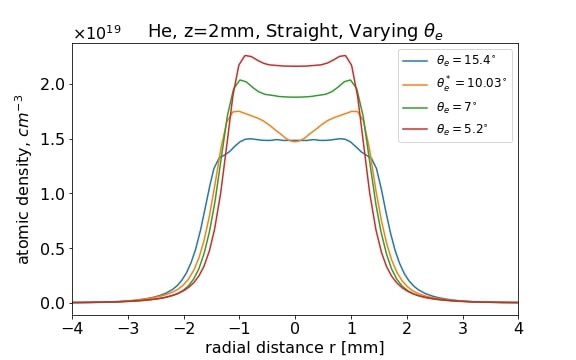}
        \caption{}
        \label{fig: z=2mmCurvesStraight}
    \end{subfigure}
    \begin{subfigure}{0.5\textwidth}
    \centering
        \includegraphics[width = \textwidth]{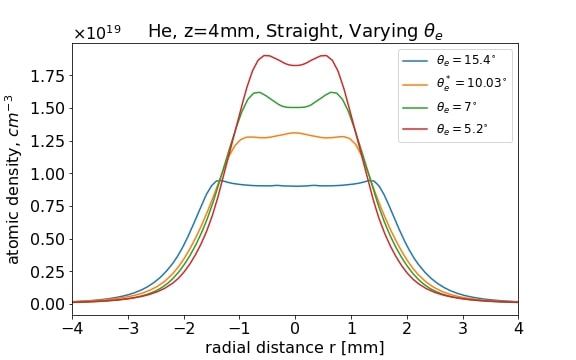}
        \caption{}
        \label{fig: z=4mmCurvesStraight}
    \end{subfigure}
    \caption{(Color) Simulation density profiles of straight nozzles with various exit half-angles, $\theta_e$, at \textbf{(a)} z = 2 mm, and \textbf{(b)} z = 4 mm. All other parameters were kept the same and all nozzles had $M_e = 5.7$. The Mach angle, $\theta_e$ = $10.03^{\circ}$, is marked with an asterisk.}
    \label{fig: straightdifftheta}
\end{figure}

For the straight nozzle, density perturbations can also be minimized to yield flat-top profiles by varying $\theta_e$, as seen in Fig. \ref{fig: straightdifftheta}, which is done by varying $l_d$ since $d^*$ and $d_e$ are held constant. Matching the nozzle exit angle $\theta_e$ to Mach angle $\sin^{-1}{(1/M_e)}$ roughly leads to a flatter profile as it minimizes the effects of the shocks in perturbing the density, agreeing with past studies on straight nozzle optimization \cite{MKrish, cgrth, Semushin}. Further refinement of $\theta_e$, specifically increasing it above the Mach angle, will lead to flatter profiles as shown in Fig. \ref{fig: straightdifftheta}. However, because the only way to control the shock features of the straight nozzle is by varying the $\theta_e$ (since $d^*$ and $d_e$ are held constant), optimizing the straight nozzle lacks the extra parameter of control in adjusting $R_c$, unlike the trumpet nozzle. This lack of curvature removes the ability to tune shock suppression in addition to using the $\theta_e$ optimization to minimize shock effects, explaining why the density profiles shown in Fig. \ref{fig: straightdifftheta} still show fluctuations near the optimal angle. Therefore, although both straight and trumpet nozzles can be optimized to create flat-top profiles, the trumpet nozzle's optimization procedure allows for additional refinement with its extra degree of freedom through its diverging section curvature.  

While the optimization of the trumpet and straight performed here was for the prescribed nozzle geometry parameters of $d^* = 0.8$ mm and $d_e = 3$ mm, these optimization processes can be done for any mm-scale axisymmmetric nozzle geometry as shown before with the straight nozzle \cite{MKrish, Semushin, cgrth}. The trumpet nozzle optimization is not specific to our geometry conditions as the radius of curvature can be similarly adjusted for a different mm-scale nozzle geometry. If the exit diameter is a different mm-scale value, one would perform this optimization process in the same manner as presented.

\section{\label{sec: discon} Discussion and Conclusion}

In this paper, we examined three different axisymmetric mm-scale nozzle geometries in the bell, straight and trumpet nozzles and investigated the effect of the curvature variation on the gas jet density field as well as shock features. The trumpet nozzle was optimized to produce consistent flat-top profiles in mm-scale distances from its exit. The trumpet optimization procedure was then compared to the straight nozzle optimization. 

The study of the diverging section curvature's effect on gas jet density was conducted in two parts. The first part involved simulating various nozzle geometries to study this effect for few mm-scale nozzles. The main result was that the nozzle curvature had noticeable impact on shock formation and the resulting gas jet density field and therefore, is an important parameter for LPA gas jet design. It was found that the trumpet geometry, like the straight and bell nozzles, could be optimized to create consistent flat-top density profiles. The trumpet $\theta_e$ optimization condition was similar to the Mach angle condition of the straight nozzle although with an added parameter of control in its radius of curvature, which allowed for additional adjustment of the nozzle's shock suppression strength. This added parameter of curvature provided an additional degree of freedom for refinement of the trumpet nozzle geometry, which allowed the trumpet nozzle to create flatter profiles at the cost of slightly wider edges compared to the straight nozzle. Although the bell nozzle has been shown in the past to be able to effectively produce flat-top profiles \cite{Lemos}, the results showed that the bell nozzle could also create a focusing effect that amplified shock diamonds near the nozzle exit, leading to density spikes, which can be exploited as a design concept for short, high density kHz LPA targets driven by few cycle laser pulses \cite{kHzLPA1, kHzLPA2}. In the second part, simulation results were verified with neutral density interferometry measurements, showing good qualitative agreement with simulation findings.

The results apply to mm-scale axisymmetric nozzles for densities between a few $10^{17}$ to $10^{20}$ cm$^{-3}$. If the exit diameter is a different value that is mm-scale, such as 5 mm, one would perform the optimization process by tuning the radius of curvature in the same manner as presented and observing the resulting density profiles. Sub-mm nozzles require additional modeling of the boundary layer as the boundary layer for such a nozzle becomes comparable to the nozzle's geometry parameters \cite{gasjetSchmid}. Furthermore, non-axisymmetric nozzles, such as rectangular slit nozzles \cite{gastargetsLorenz}, have different underlying physical processes for shock formation and flow. Thus, the results cannot be conclusively generalized to sub-mm nozzles nor non-axisymmetric nozzles and require further examination.

The manufactured trumpet nozzle will be applied in ongoing LPA-based MeV Thomson photon source experiments, leading the way to a compact, affordable and narrow bandwidth x-ray source \cite{thomsonmps}. Future work will focus on designing nozzles with tailored density profiles, e.g., to integrate injection, acceleration and deceleration in one jet or to optimize betatron radiation with multiple sections of varying density \cite{tomkus, phuoc, betaPhysLett}.

\section*{Supplementary Material}
See supplementary material for information on manufacturing costs as well as additional figures referenced in manuscript. 

\section*{Acknowledgements}
This work is supported by the U.S. Department of Energy, NNSA DNN R\&D and SC HEP under Contract No.
DE-AC02-05CH11231. This material is also based upon work supported by the Department of
Energy National Nuclear Security Administration through the Nuclear Science and
Security Consortium under Award Number(s) DE-NA0003180 and/or DE-NA0000979.

The authors gratefully acknowledge the technical support from Zachary Eisentraut and Tyler Sipla.

\section*{Data Availability}
Raw data was generated at Lawrence Berkeley National
Laboratory. The data that support the findings of this study are
available from the corresponding author upon reasonable request.

\nocite{*}

\providecommand{\noopsort}[1]{}\providecommand{\singleletter}[1]{#1}%

\end{document}